\documentstyle[twocolumn,seceq]{jpsj}
\input epsf.tex

\def\runtitle{Crossover between Fermi Liquid and non-Fermi Liquid in Orbitally Degenerate Kondo Systems}
\def\runauthor{Hiroaki {\sc Kusunose} and Yoshio {\sc Kuramoto}}

\title
{Crossover between Fermi Liquid and non-Fermi Liquid in Orbitally Degenerate Kondo Systems}

\author
{
Hiroaki {\sc Kusunose}\footnote{E-mail: kusu@cmpt01.phys.tohoku.ac.jp}
and Yoshio {\sc Kuramoto}
}

\inst
{
Department of Physics, Tohoku University, Sendai, 980-8578
}

\recdate
{
July 7, 1999}

\abst
{
Entanglement of spin and orbital Kondo effect is investigated on the basis
of a Kondo-type exchange model with twofold orbital degeneracy.
By using Wilson's numerical renormalization-group method, we examine dynamical and thermal properties respecting the difference in time-reversal property of multipole operators.
In the presence of particle-hole symmetry, the model has a new non-Fermi-liquid fixed point with a fractional entropy.
The spectral intensity of the quadrupole susceptibility diverges in the zero-frequency limit, while the dipole susceptibility shows a Fermi-liquid-like behavior.
This is understood by mapping to the two-channel Kondo model, in which the dipole moment is mapped onto the operators with the scaling dimension $\Delta_m=1$, while the quadrupole moment onto the operators with another scaling dimension $\Delta_e=1/2$.
Even for a fairly particle-hole asymmetric case with the Fermi-liquid ground state, the non-Fermi-liquid behavior has significant influences in electric and thermal properties.
}

\kword
{
orbital degeneracy, time-reversal property, multipole, numerical
renormalization-group method, multi-channel Kondo effect, non-Fermi liquid, Ce$_x$La$_{1-x}$B$_6$, La$_{1-x}$Sr$_x$MnO$_3$
}

\begin{document}
\sloppy
\maketitle

%
%
\section{Introduction}
Orbital dynamics has revived interest because of new experimental progress in heavy-fermion systems such as Ce$_x$La$_{1-x}$B$_6$\cite{nakamura,tayama,hiroi,suzuki} and transition metal oxides such as La$_{1-x}$Sr$_x$MnO$_3$\cite{ramirez,mori,murakami}.
Much attention has been attracted to physical origin of anomalous property in magnetic, elastic and transport properties in these systems.
Multipole fluctuations seem to play an essential role
in realizing these properties.\cite{uimin,fukushima,motome,maezono}

In the presence of orbital degeneracy, orbital multipoles as well as spin ones constitute observables.\cite{kugel,schmitt,ohkawa,shiina}
The simplest model to describe the combined spin and orbital degrees of freedom is the SU(4) symmetric spin-orbit model.
As a consequence of the entanglement of spin and orbital multipoles,
fluctuations with the four-site periodicity govern the low-energy physics in one dimension.\cite{sutherland,affleck,yamashita,li,frischmuth}
On the other hand, an on-site fluctuation with conduction electrons is examined by the SU(4) Coqblin-Schrieffer (CS) model.
The Kondo effect based on this model is well understood at present.\cite{rajan,sakai}
In these models, however, the difference between the spin and the orbital multipoles is hidden in the excessively high symmetry.

In order to take into account the difference, we have proposed\cite{kusunose1,kusunose2}
that the time-reversal property of multipole operators is essential, since the distinction between magnetic and electric characters should be important in interpreting experimental results, including dilution effects.
For instance, the dipole operator couples with external magnetic field, while the quadrupole operator interacts with distortion of the lattice.

In the impurity problem, the CS model was extended toward the realistic atomic-level structure by Hirst.\cite{hirst1,hirst2}
The model was investigated by the perturbational renormalization-group analysis.\cite{pavarini}
In such context, the spin-channel Kondo model was investigated by several authors\cite{pang,ye,kuramoto}, in which the impurity possesses both spin and channel degrees of freedom.
The present authors examined more general exchange interaction in accordance with the point-group symmetry.\cite{kusunose1}
A scaling analysis was adopted to investigate the time-reversal difference of the multipole operators in the exchange model with arbitrary number of orbital degeneracies.\cite{kusunose2}
We found a new class of non-Fermi-liquid fixed point in the presence of the particle-hole symmetry.
It was argued that the new fixed point exhibits anomalous behaviors only in electric and thermal properties.

The purpose of this paper is to perform quantitative investigation of the model by a non-perturbative method.
Namely, we use Wilson's numerical renormalization-group (NRG) method\cite{wilson} to derive dynamical and thermal properties.\cite{krishna,sakai,costi}
In the presence of the particle-hole symmetry, the system in fact has a non-Fermi-liquid fixed point with a fractional entropy.
Near the new fixed point, the spectral intensity of the quadrupole susceptibility diverges in the zero-frequency limit, while the dipole susceptibility shows a Fermi-liquid-like behavior.
The quantative investigation reveals that for a fairly particle-hole asymmetric case the non-Fermi-liquid behavior has a significant influence in electric and thermal properties in the crossover process to the Fermi-liquid ground state.

The organization of the paper is as follows.
In the following section, the exchange model with twofold orbital degeneracy is explained. Then, some preliminaries to the NRG calculation are given.
The results of dynamical and thermal properties are given in \S 3.
The nature of the non-Fermi-liquid fixed point is discussed in \S 4.
The final section summarizes the paper.

%
%
\section{Model}
We begin with a Kondo-type exchange model with twofold orbital degeneracy.
Let us write the local multiplet as $|\ell\sigma\rangle$, where $\sigma$ represents a Kramers pair $\uparrow$, $\downarrow$ with arbitrary strong spin-orbit interaction and the crystalline electric field effect.
The orbital pair $\ell=1,2$ can be represented by a pseudospin $\tau^z = \pm 1$.
The interaction Hamiltonian is then given by
\begin{eqnarray}
&&H_{\rm ex}=
\frac{1}{4}J_m\left[{\mib \sigma}_c\cdot{\mib \sigma}_f+\tau^y_c\tau^y_f
+(\tau^x_c\tau^x_f+\tau^z_c\tau^z_f){\mib \sigma}_c\cdot{\mib \sigma}_f
\right]
\nonumber\\&&\mbox{\hspace{1cm}}
+\frac{1}{4}J_e\left[(\tau^x_c\tau^x_f+\tau^z_c\tau^z_f)+\tau^y_c\tau^y_f{\mib \sigma}_c\cdot{\mib \sigma}_f\right],
\label{previous-H}
\end{eqnarray}
where the operators with suffix $c$ ($f$) denote conduction- ($f$-) electron operators at the impurity site.
The anisotropy in $\tau$-space results from the odd character of $\tau^y$ under time-reversal operation.\cite{kusunose1}
Here two different coupling constants $J_m$ and $J_e$ have been used to characterize interactions of magnetic and electric origins.
The SU(4) CS model can be reproduced by putting $J_m=J_e$.
The generalization to the realistic point-group symmetry is straightforward.\cite{kusunose1}

In order to analyze the time-reversal property conveniently, we express the local operators in alternative basis, i.e., azimuthal component $m$ of the fictitious spin $j=3/2$:
\begin{equation}
(1\uparrow,2\downarrow,2\uparrow,1\downarrow)\to(+3/2,+1/2,-1/2,-3/2).
\end{equation}
In the new expression, one can easily see that the quadrupole operator connects
one orbital with another, while the dipole and the octupole operators couple
different spins.
The former has an even property under time-reversal operation while the latter has an odd property.

The interaction among operators between $f$ and conduction electrons is expressed in the scalar form of the spherical tensors\cite{kusunose2}:
\begin{equation}
H_{\rm ex}=\sum_{p=1}^{3}J_p\sum_{q=-p}^p
c^\dagger {\sf T}^{(p)}_q c\; |f\rangle [{\sf T}^{(p)}_q]^\dagger\langle f|
\label{H_int}.
\end{equation}
It is noted that the spherical symmetry SU(2) of the model is adopted simply for technical convenience.
Here ${\sf T}^{(p)}_q$ is $4\times4$ matrix representation of the $q$-th component of the spherical tensor operator with rank $p$.\cite{landau}
These are expressed explicitly in terms of the Clebsch-Gordan coefficient,
\begin{equation}
({\sf T}^{(p)})_{mm'}=\frac{(-{\rm i})^p}{2}(2p+1)^{1/2}\langle 3/2m|pq;3/2m'\rangle,
\end{equation}
and satisfies the orthonormality,
\begin{equation}
{\rm Tr}({\sf T}^{(p)}_q[{\sf T}^{(p')}_{q'}]^\dagger)
=\delta_{pp'}\delta_{qq'}.
\end{equation}
Accordingly, the conduction and $f$ operators are represented in the symbolic vector notations in eq. (\ref{H_int}).
If one considers the time-reversal difference in eq. (\ref{H_int}), i.e., $J_m\equiv J_1=J_3$ and $J_e\equiv J_2$, the Hamiltonian becomes equivalent to eq. (\ref{previous-H}).

To apply the NRG method to the present model\cite{wilson,krishna,sakai,costi}, the Hamiltonian has to be cast into the semi-infinite linear chain form.  With the discretization parameter $\Lambda$, it is given by
\begin{eqnarray}
&&H_N=\Lambda^{(N-1)/2}\left[H_k+H_{\rm ex}\right],
\label{nrgham}
\\&&\mbox{\hspace{5mm}}
H_k=\sum_m\sum_{n=0}^{N-1}\Lambda^{-n/2}(c^\dagger_{nm}c_{n+1m}+{\rm h.c.}),
\\&&\mbox{\hspace{5mm}}
H_{\rm ex}=\sum_{p=1}^{3}J_p\sum_{q=-p}^p
t^{(p)}_q\; [T^{(p)}_q]^\dagger,
\end{eqnarray}
where the ``site'' in the linear chain is denoted by the label $n$.
The localized tensor operators for conduction and $f$ electrons are defined as
\begin{eqnarray}
&&t^{(p)}_q=\sum_{mm'}c^\dagger_{0m}({\sf T}^{(p)}_q)_{mm'}c_{0m'},
\\&&
T^{(p)}_q=\sum_{MM'}|M\rangle({\sf T}^{(p)}_q)_{MM'}\langle M'|.
\end{eqnarray}

As conserved quantum numbers, the spherical symmetry allows us to use the total spin in addition to the total charge number measured from the half-filled state.
These quantum numbers are defined as the eigenstates of the following
operators:
\begin{eqnarray}
&&
S^\alpha_N=\sum_{n=0}^N\sum_{mm'}c^\dagger_{nm}{\sf S}^\alpha_{mm'}c_{nm'}
\nonumber\\&&\mbox{\hspace{2cm}}
+\sum_{MM'}|M\rangle {\sf S}^\alpha_{MM'}\langle M'|,
\\&&
Q_N=\sum_m\sum_{n=0}^N\left(c^\dagger_{nm}c_{nm}-\frac{1}{2}\right),
\end{eqnarray}
where the dipole moments are related to the spherical tensors of rank 1,
\begin{equation}
\sqrt{5}{\sf T}^{(1)}_0={\rm i}\;{\sf S}^z,\;\;\;\;\sqrt{5}{\sf T}^{(1)}_{\pm 1}=\mp\frac{\rm i}{\sqrt{2}}({\sf S}^x\pm i{\sf S}^y).
\end{equation}

%
%
\section{Results}
In the preset work, we have used $\Lambda=3$ and about 900 states have been retained at each step of the renormalization-group transformation.
The transformation is given by
\begin{equation}
H_{N+1}=\Lambda^{1/2}H_N+\sum_{m}(c^\dagger_{Nm}c_{N+1m}+{\rm h.c.}).
\end{equation}
We take the unit of energy as $D(1+\Lambda^{-1})/2$ of the modified bandwidth due to discretization.

\subsection{Characteristic energies}
Let us first discuss the characteristic energies in the system.
According to the scaling analysis, there are two characteristic energies corresponding to magnetic and electric responses.\cite{kusunose2}
In order to define the characteristic energies explicitly, one uses dynamical
susceptibilities for each rank of the localized tensors,
\begin{equation}
\chi_p(\omega)={\rm i}\int_0^\infty {\rm d}t \exp({\rm i}\omega t)\langle[T^{(p)}_q(t),(T^{(p)}_q)^\dagger]\rangle,
\end{equation}
where the parenthesis denotes the average over the ground state.
Notice that $\chi_p$ is independent of the component index $q$ due to the spherical symmetry.
Since the condition $J_1=J_3$ ensures $\chi_1=\chi_3$,
$\chi_1(\omega)$ is termed in the following the magnetic susceptibility $\chi_m(\omega)$,
and $\chi_2(\omega)$  the electric susceptibility $\chi_e(\omega)$.

As in the standard Kondo system, the characteristic energy for the magnetic response is defined by the inverse of the static susceptibility, i.e., $T_m=1/\chi'_m(0)$.
Since the NRG method provides us with the spectral intensity, the static susceptibility can be computed by means of the Kramers-Kronig relation.
On the other hand, the static quadrupolar susceptibility diverges at $J_e=0$.
Then, we have to choose alternative definition for the characteristic energy.
We use $T_e=\alpha\omega_{\rm max}$, where $\omega_{\rm max}$ is the peak position of the spectral intensity.
Here we have introduced the coefficient $\alpha$ so as to make the two characteristic energies be equal in the case of $J_m=J_e$.
The numerical calculation gives $\alpha=3.6952$.

\begin{figure}
\vspace{1mm}
\begin{center}
\epsfxsize=8cm \epsfbox{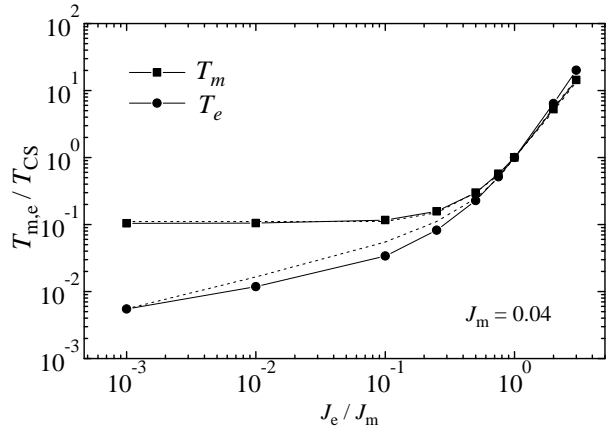}
\end{center}
\caption{The magnetic and electric characteristic energies $T_m$ (square) and $T_e$ (circle) as a function of the ratio of the couplings. The values are scaled by the characteristic energy of the Coqblin-Schrieffer model $T_{\rm CS}$.
The dashed lines are the estimated results by the scaling analysis.
}
\label{figtk}
\end{figure}

The characteristic energies $T_m$ and $T_e$ is shown in Fig. \ref{figtk}
as a function of $J_e/J_m$.
The energies are scaled by the characteristic energy $T_{\rm CS}$ of the CS model.
The dashed lines show the results by the scaling theory at the two-loop level.
The $T_K$'s are defined by the condition $J_{\rm eff} = 0.5$ where the energy scale of interest becomes $T_K$.
The numerical results show good agreement with the scaling theory.
The difference between the two characteristic energies is conspicuous for $J_e/J_m<0.3$,
while the difference for $J_e/J_m>1$ remains small.
The saturation of $T_m$ as $J_e/J_m\to 0$ indicates that the magnetic property is not affected by the electric one for small $J_e/J_m$.

\subsection{Dynamical susceptibilities}
\begin{figure}
\vspace{1mm}
\begin{center}
\epsfxsize=8cm \epsfbox{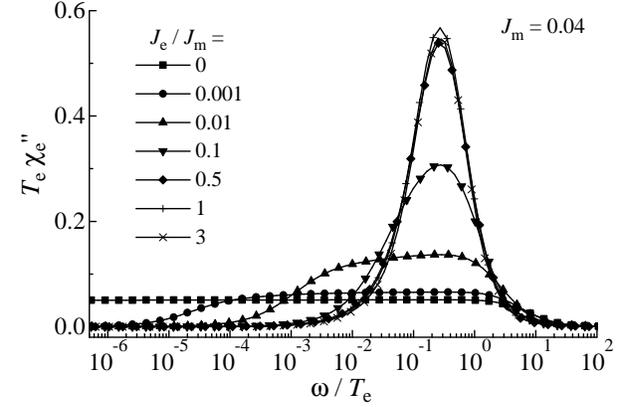}
\end{center}
\caption{The spectral intensity of the electric excitation related to the quadrupole response. The zero-frequency limit of the intensity has a finite value in the case $J_e=0$.}
\label{figchie}
\end{figure}

We show first the results of the electric response.
The spectral intensity $\chi''_e$ of the electric excitation is shown in Fig. \ref{figchie}.
The dominant intensity, which has frequencies $\omega/T_e\sim 0.3$ for $J_e/J_m > 0.1$, moves to the low-frequency region as $J_e/J_m$ decreases.
In the case of $J_e/J_m=0$,  the intensity remains finite at zero frequency.
Thus one can conclude that  the quadrupole response exhibits a non-Fermi-liquid behavior with divergent static susceptibility.

\begin{figure}
\vspace{1mm}
\begin{center}
\epsfxsize=8cm \epsfbox{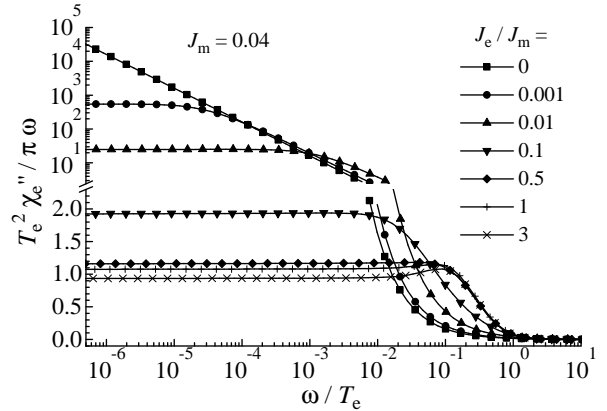}
\end{center}
\caption{A plot of $\chi''_e/\omega$ as a function of $\omega$ for various $J_e/J_m$. The low-frequency intensity is extremely enhanced as $J_e/J_m$ decreases, reflecting divergent aspect in the case $J_e=0$. Note that both linear and logarithmic scales are used in azimuthal axis.
}
\label{figchiwe}
\end{figure}

A plot of $\chi''_e/\omega$ as a function of $\omega/T_e$ is shown in Fig. \ref{figchiwe}.
Reflecting divergent behavior in the case $J_e=0$, the intensity in the low-frequency region is extremely enhanced as $J_e/J_m$ decreases.
However, the zero-frequency limit of the intensity saturates to a finite value
as long as the electric coupling is nonzero.
Although the system eventually flows to the Fermi-liquid fixed point with finite $J_e$,
the quadrupole response cannot be characterized by an universal function of $\omega/T_e$.

\begin{figure}
\vspace{1mm}
\begin{center}
\epsfxsize=8cm \epsfbox{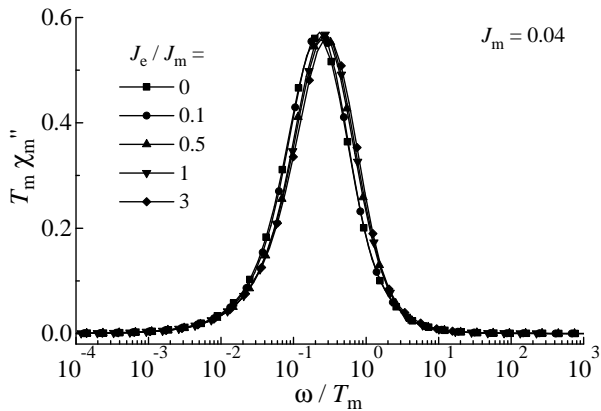}
\end{center}
\caption{The spectral intensity of the magnetic excitation related to the dynamical response of the dipole and the octupole operators.
The results for various ratio of the couplings $J_e/J_m$ are scaled by the function of $\omega/T_m$.}
\label{figchim}
\end{figure}

The magnetic spectral intensity $\chi''_m(\omega)$ for various $J_e/J_m$ is shown in Fig. \ref{figchim}.
In contrast to the electric response, the results for any value of $J_e/J_m$ are scaled very well by a universal function of $\omega/T_m$ with the peak at $\omega/T_m\sim 0.3$.
We should emphasize that the zero frequency limit of $\chi''_m$ converges to zero even for $J_e=0$.

\begin{figure}
\vspace{1mm}
\begin{center}
\epsfxsize=8cm \epsfbox{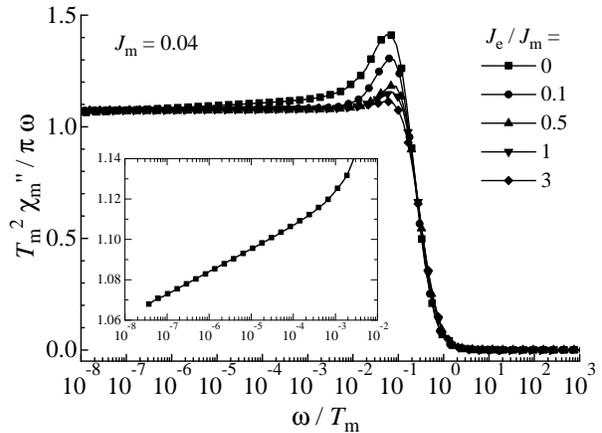}
\end{center}
\caption{A plot of $\chi''_m/\omega$ as a function of $\omega$ for various $J_e/J_m$. The logarithmic correction exists in the case of $J_e=0$ (inset).}
\label{figchiwm}
\end{figure}

A plot of $\chi''_m/\omega$ as a function of $\omega/T_m$ is shown in Fig. \ref{figchiwm}.
The peak at $\omega/T_m\sim 0.1$ is enhanced as $J_e/J_m$ decreases.
The zero-frequency limit satisfies the Korringa-Shiba relation,\cite{ksrel}
\begin{equation}
\lim_{\omega\to0}T_m^2\chi''_m/\pi\omega=1,
\end{equation}
within the numerical accuracy of about 10\%.
It means that the magnetic response satisfies the Fermi-liquid relation for any ratio of the couplings $J_e/J_m$, including the case $J_e=0$.
The reason for this will be explained in the next section.
It should be stressed that for $J_e=0$ a logarithmic correction, $c\log\omega$, is present as shown in the inset of Fig. \ref{figchiwm}.
Even though the correction is rather small ($c\sim 10^{-2}$), this is  the only trace of the non-Fermi-liquid fixed point in the magnetic response.
The magnetic response can be scaled mostly by the universal function of $\omega/T_m$, apart from the difference in the peak height.

\subsection{$T$-matrix}
The $T$-matrix of conduction electrons can be computed\cite{brenig} by the following correlation function,
\begin{equation}
T(\omega)=-{\rm i}\int_0^\infty {\rm d}t {\rm e}^{{\rm i}\omega t}\langle \{j_m(t),j_m^\dagger\} \rangle,
\end{equation}
where the current operator $j_m$ is given by
\begin{eqnarray}
j_m&=&\left[c_{km},H_{\rm ex}\right]\nonumber\\
&=&\frac{1}{\sqrt{2}} \sum_p J_p \sum_{q}\sum_{m'}({\sf T}^{(p)}_q)_{mm'}c_{0m'}[T^{(p)}_q]^\dagger.
\label{currentop}
\end{eqnarray}
Note that the current operator is independent of the wave vector.

\begin{figure}
\vspace{1mm}
\begin{center}
\epsfxsize=8cm \epsfbox{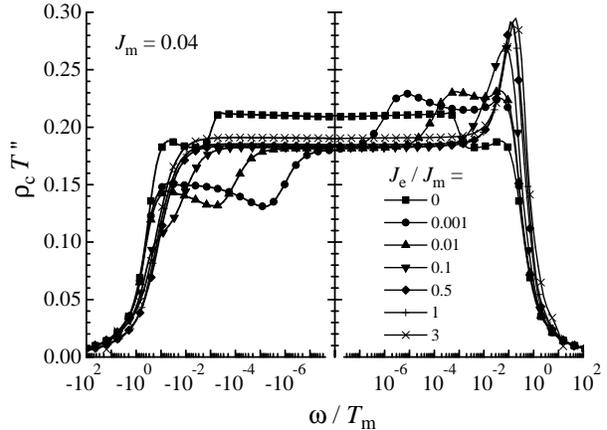}
\end{center}
\caption{The imaginary part of the $T$-matrix. The energy $\omega$ is scaled by the magnetic characteristic energy $T_m$.}
\label{figtw}
\end{figure}

The imaginary part of the $T$-matrix is shown in Fig. \ref{figtw}.
In the CS limit, the single peak at $\omega/T_m\sim 0.1$ is consistent with the previous works for the CS model\cite{rajan} and the corresponding Anderson model.\cite{sakai}
As $J_e/J_m$ decreases, the single peak extends down to the lower frequency region, and simultaneously the intensity below the Fermi level decreases.
The flat region straddling the Fermi level shrinks to extent which correlates directly with the crossover energy to the Fermi-liquid fixed point.
The crossover energy can easily be estimated in results of the entropy and the specific heat.
This suggests that the conventional $T^2$ behavior of the resistivity can be seen only in lower temperature.
If the coupling constants are small enough as compared with the bandwidth, the zero-frequency limit satisfies the Friedel sum-rule.

It is noted that for $J_e/J_m<0.01$, $T''$ relative to its value at the Fermi level seems antisymmetric as a function of $\omega$.
In the limit of $J_e=0$, on the other hand, the $T$-matrix is a symmetric function of $\omega$ due to the particle-hole symmetry.
Thus, the structure of the $T$-matrix must change discontinuously at $J_e=0$.

In the case of $J_e=0$, the residual resistivity associated with the zero-frequency limit of the $T$-matrix is larger than that in the case $J_e/J_m\ne 0$.
The constant behavior in the $T$-matrix is somewhat surprising in spite of the singular behavior in the quadrupole susceptibility.
Note that in the case of the conventional two-channel model\cite{cox}, the self-energy of the conduction electron\cite{ludwig,affleck2} and the single-particle excitation of the localized electron\cite{ssuzuki} behaves as $|\omega|^{1/2}$ due to the leading irrelevant operator.
In the present case, however, the singular contribution of the quadrupole moment to the current opeartor vanishes because $J_e=0$ in eq. (\ref{currentop}).

\subsection{Entropy and specific heat}
\begin{figure}
\vspace{1mm}
\begin{center}
\epsfxsize=8cm \epsfbox{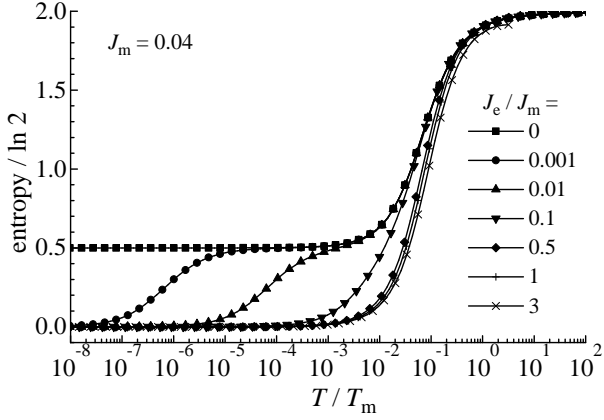}
\end{center}
\caption{The temperature dependence of the impurity entropy. The temperature is scaled by the magnetic characteristic energy.}
\label{figentropy}
\end{figure}

The impurity contribution to the entropy is shown in Fig. \ref{figentropy}.
In the case of $J_e=0$, the residual entropy has the fractional value, $\ln\sqrt{2}$, which is the same as the value in the two-channel Kondo model.
This implies that the low-energy spectrum is determined by the same fixed-point Hamiltonian as the two-channel Kondo model.
Once the coupling $J_e$ is switched on, the residual entropy vanishes.
However, the entropy takes the constant value over some temperature range
if $J_e/J_m<0.01$.
The smaller $J_e/J_m$ gives the wider range to observe the fractional entropy.

\begin{figure}
\vspace{1mm}
\begin{center}
\epsfxsize=8cm \epsfbox{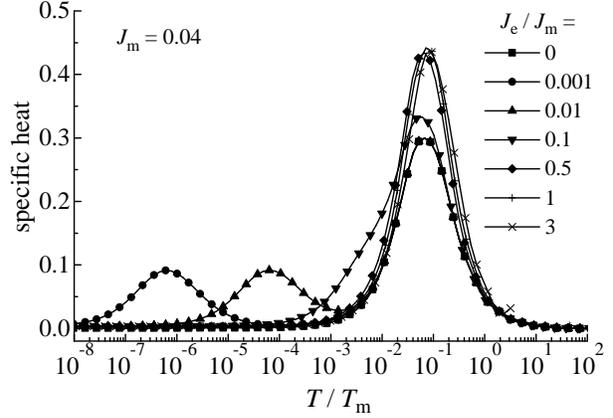}
\end{center}
\caption{The impurity specific heat.}
\label{figspecific}
\end{figure}

The impurity contribution to the specific heat is shown in Fig. \ref{figspecific}.
The double-peak structure in the specific heat reflects the release of the entropy in two steps.
Two peaks can be separable for $J_e/J_m<0.01$.

\section{Non-Fermi-liquid fixed point}
\begin{figure}
\vspace{1mm}
\begin{center}
\epsfxsize=8cm \epsfbox{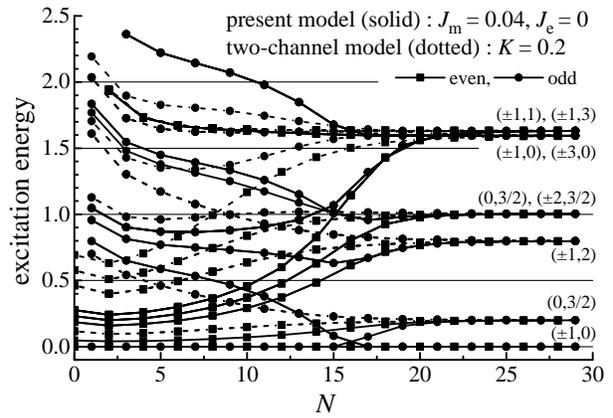}
\end{center}
\caption{The flow diagram of the excited energy spectrum for the present model (solid line) and the two-channel Kondo model (dotted line).}
\label{figflow}
\end{figure}

Now, we turn to discuss a new class of non-Fermi-liquid fixed point in the presence of the particle-hole symmetry, i.e., $J_e/J_m=0$.
The flow diagram of the energy spectrum are shown in Fig. \ref{figflow}, as the renormalization step $N$ proceeds.

The solid lines show the result of the present model, eq. (\ref{nrgham}), while the dotted lines correspond to the two-channel Kondo model\cite{cox} with the exchange interaction,
\begin{equation}
H_{\rm TC}=K\sum_{\ell}\sum_{\sigma\sigma'}c^\dagger_{0\ell\sigma}{\mib\sigma}_{\sigma\sigma'}c_{0\ell\sigma'}\cdot{\mib S},
\end{equation}
where the $\ell$ and $\sigma$ specify the flavor (channel) and the spin degrees of freedom, respectively.
${\mib S}$ is the localized spin of $S=1/2$.
The labels referring to each line denote the conserved quantum numbers of the present model $(Q,J)$.

It should be emphasized that the energy spectrum of the present model is identical to
the two-channel Kondo model near the fixed point.
The correspondence of the quantum numbers are listed in Table \ref{table1}, where the total charge $Q$, the flavor-spin $F$ and the spin $S$, have been used for the two-channel model.

\begin{table}
\caption{The correspondence of the quantum numbers between (a) the present model and (b) the two-channel model.}
\label{table1}
\begin{tabular}{@{\hspace{\tabcolsep}\extracolsep{\fill}}ll} \hline
(a) ($Q,S$) & (b) ($Q,F,S$) \\ \hline
$(\pm 1,0)$ & $(0,0,1/2)$ \\
$(0,3/2)$ & $(\pm 1,1/2,0)$ \\
$(\pm 1,2)$ & $(0,1,1/2)$, $(\pm 2,0,1/2)$ \\
$(0,3/2)$, $(\pm 2,3/2)$ & $(\pm 1,1/2,1)$ \\
$(\pm 1,0)$, $(\pm 3,0)$ & $(0,0,3/2)$ \\
$(\pm 1,1)$, $(\pm 1,3)$ & $(0,0,1/2)$, $(0,1,1/2)$, $(\pm 2,1,1/2)$ \\
\hline
\end{tabular}
\end{table}

The low-energy states in two models have the same degeneracy, and the spectrum of charge excitation in the present model has a one-to-one correspondence to the spectrum of spin excitation in the two-channel model.
Thus, the effective interaction of the present model is equivalent to that of the two-channel Kondo model.
The spin degrees of freedom in the two-channel Kondo model is attributed to the charge ones (isospin) in the present model.\cite{kusunose2}

We discuss now critical behavior of the dynamical susceptibilities based on the two-channel Kondo model.
Let us assume that all matrix elements $|\langle 0|T^{(p)}_q|\alpha\rangle|^2$ decrease as $\Lambda^{-\Delta_p N}$.
$\Delta_p$ is the scaling dimension of a boundary operator in the terminology of the conformal field theory.
Then, one obtains
\begin{equation}
\chi''_p(\omega)\sim{\rm Im}\sum_\alpha\frac{|\langle 0|T^{(p)}_q|\alpha\rangle|^2}{\omega-E_\alpha+{\rm i}\delta} \sim\Lambda^{-(2\Delta_p-1)N/2} \sim\omega^{2\Delta_p-1},
\end{equation}
where we have used the fact the energy scale varies as $\Lambda^{-N/2}$ in the renormalization evolution.
Indeed, the numerical calculation shows the Fermi-liquid exponent $\Delta_m=1.00134$ for the magnetic susceptibility, while the non-Fermi-liquid one $\Delta_e=0.500502$ for the electric one.

Corresponding to the magnetic susceptibility in the present model, the matrix elements are finite only with the excited states
$\alpha=(Q,F,S)=(0,0,1/2)$, $(0,1,1/2)$, $(\pm2,1,1/2)$ in the two channel model.
Then, the boundary operators are characterized by the quantum numbers $(0,0,0)$, $(0,1,0)$ and $(\pm2,1,0)$, which correspond to the density, the flavor-spin density and the flavor-triplet pair operators, respectively.
These operators are classified together by the $SO(5)$ representation with its dimensionality $d=10$.
In fact, they have the scaling dimension $\Delta_m=1$ at the fixed point of the two-channel Kondo model\cite{ludwig2}.
Similar argument leads to the conclusion that the boundary operators corresponding to the electric susceptibility belong to $d=5$ with $\Delta_e=1/2$, which contains the flavor-spin density operator, $(0,1,0)$ and the flavor-spin singlet pair operator, $(\pm2,0,0)$.

%
%
\section{Summary}
In summary, we have investigated quantitative aspects of dynamical and thermal properties on spin and orbital Kondo effect in orbitally degenerate systems.
In order to distinguish the spin and orbital degrees of freedoms, we have paid attention to the time-reversal difference of the multipole operators.
The difference is taken into account by distinguishing between the coupling constant $J_m$ for the dipole as well as the octupole operators, and another constant $J_e$ for the quadrupole operator.

By using Wilson's NRG method, we have identified a new class of the non-Fermi-liquid fixed point in the presence of the particle-hole symmetry.
The spectral intensity of the quadrupole susceptibility diverges in the zero-frequency limit as expected.
By contrast, the dipole susceptibility shows a Fermi-liquid-like behavior, even though the fixed point has the non-Fermi-liquid excitation spectrum.
This is understood by mapping onto the two-channel Kondo model.
Namely in terms of the conformal field theory, correspondence with the two-channel Kondo model has been made as follows:
the dipole moment in the present model is mapped onto three operators with the dimension $\Delta_m=1$, the density, the flavor-spin density and the flavor-singlet pair operators.
On the other hand, the quadrupole moment pretends as two $\Delta_e=1/2$ irrelevant opeartors, the flavor-spin density and the flavor-spin singlet pair ones.
Thus, the Fermi-liquid-like behavior appears only in the magnetic response.

Even in realistic situations with a rather large particle-hole asymmetry, the anomalous behavior discussed in this paper should appear in electric and thermal properties in the temperature range accessible by experiments.

%
%
\section*{Acknowledgements}
We would like to thank Prof. Osamu Sakai, Dr. Yukihiro Shimizu and
Dr. Shunya Suzuki for discussion on details of the NRG computation.
Thanks are also due to  Prof. Masaki Oshikawa for fruitful discussion.
This work was supported by a Grant-in-Aid for the encouragement of Young
Scientists from the Ministry of Education, Science, Sports and Culture of
Japan.

%
%


\begin{thebibliography}{99}
\bibitem{nakamura} S. Nakamura, O. Suzuki, T. Goto, S. Sakatsume, T. Matsumura and S. Kunii: J. Phys. Soc. Jpn. {\bf 66} (1997) 552.
\bibitem{tayama} T. Tayama, T. Sakakibara, K. Tenya, H. Amitsuka and S. Kunii: J. Phys. Soc. Jpn. {\bf 66} (1997) 2268.
\bibitem{hiroi} M. Hiroi, S. Kobayashi, M. Sera, N. Kobayashi and S. Kunii: J. Phys. Soc. Jpn. {\bf 67} (1998) 53.
\bibitem{suzuki} O. Suzuki, T. Goto, S. Nakamura, T. Matsumura and S. Kunii: J. Phys. Soc. Jpn. {\bf 67} (1998) 4243.
\bibitem{ramirez} A. P. Ramirez: J. Phys. {\bf C9} (1997) 8171.
\bibitem{mori} S. Mori, C. H. Chen and S.-W. Cheong: Nature {\bf 392} (1998) 473.
\bibitem{murakami} Y. Murakami, H. Kawada, H. Kawata, M. Tanaka, T. Arima, Y. Moritomo, and Y. Tokura: Phys. Rev. Lett. {\bf 80} (1998) 1932.
\bibitem{uimin} G. Uimin, Y. Kuramoto and N. Fukushima: Solid State Commun. {\bf 97} (1996) 595.
\bibitem{fukushima} N. Fukushima and Y. Kuramoto: J. Phys. Soc. Jpn. {\bf 67} (1998) 2460.
\bibitem{motome} Y. Motome and M. Imada: preprint (cond-mat/9903183).
\bibitem{maezono} R. Maezono, S. Murakami, N. Nagaosa, S. Ishihara, M. Yamanaka and H. C. Lee: preprint (cond-mat/9905072).
\bibitem{kugel} K. I. Kugel and D. I. Khomskii: Sov. Phys. JETP {\bf
37} (1973) 725.
\bibitem{schmitt} D. Schmitt and P. M. Levy: J. Magn. \& Magn. Mater.
{\bf 49} (1985) 15.
\bibitem{ohkawa} F. J. Ohkawa: J. Phys. Soc. Jpn. {\bf 54} (1985) 3909.
\bibitem{shiina} R. Shiina, H. Shiba and P. Thalmeier: J. Phys. Soc. Jpn. {\bf 66} (1997) 1741.
\bibitem{sutherland} B. Sutherland: Phys. Rev. {\bf B12} (1975) 3795.
\bibitem{affleck} I. Affleck: Nucl. Phys. {\bf B265} (1986) 409.
\bibitem{yamashita} Y. Yamashita, N. Shibata and K. Ueda: Phys. Rev.
{\bf B58} (1998) 9114.
\bibitem{li} Y. Q. Li, M. Ma, D. N. Shi and F. C. Zhang: Phys. Rev.
Lett. {\bf 81} (1998) 3527.
\bibitem{frischmuth} B. Frischmuth, F. Mila and M. Troyer: Phys. Rev.
Lett. {\bf 82} (1999) 835.
\bibitem{rajan} V. T. Rajan: Phys. Rev. Lett. {\bf 51} (1983) 308.
\bibitem{sakai} O. Sakai, Y. Shimizu and T. Kasuya: J. Phys. Soc. Jpn. {\bf 58} (1989) 3666.
\bibitem{kusunose1} H. Kusunose and Y. Kuramoto: Phys. Rev. {\bf B59} (1999) 1902.
\bibitem{kusunose2} H. Kusunose and Y. Kuramoto: J. Phys. Soc. Jpn. {\bf 68} (1999) 1805.
\bibitem{hirst1} L. L. Hirst: Z. Phys. {\bf 244} (1971) 230
\bibitem{hirst2} L. L. Hirst: Adv. Phys. {\bf 27} (1978) 231.
\bibitem{pavarini} E. Pavarini and L. C. Andreani: Phys. Rev. {\bf B56} (1997) 5073.
\bibitem{pang} H. Pang: Phys. Rev. Lett. {\bf 73} (1994) 2736.
\bibitem{ye} J. Ye: Phys. Rev. {\bf B56} (1997) R489; Phys. Rev. Lett. {\bf 79} (1997) 1385.
\bibitem{kuramoto} Y. Kuramoto: Eur. Phys. J. {\bf B5} (1998) 457.
\bibitem{wilson} K. G. Wilson: Rev. Mod. Phys. {\bf 47} (1975) 773.
\bibitem{krishna} H. R. Krishna-murthy, J. W. Wilkins and K. G. Wilson: Phys. Rev. {\bf B21} (1980) 1003; 1044.
\bibitem{costi} T. A. Costi, A. C. Hewson and V. Zlati\'c: J. Phys. Condens. Matter {\bf 6} (1994) 2519.
\bibitem{landau} L. D. Landau and E. M. Lifshitz: {\it Quantum Mechanics} (Pergamon Press, London, 1958).
\bibitem{ksrel} This expression differ from the usual form in the numerical constants. This is only because the definitions of the multipole tensors and the characteristic energies.
\bibitem{brenig} W. Brenig and J. Zittartz: {\it Magnetism}, ed. H. Suhl (Academic Press, 1973) Vol. 5, p. 185.
\bibitem{cox} For a review, see D. L. Cox and A. Zawadowski: Adv. Phys. {\bf 47} (1998) 599, and references therein.
\bibitem{ludwig} A. W. W. Ludwig and I. Affleck: Phys. Rev. Lett. {\bf 67} (1991) 3160.
\bibitem{affleck2} I. Affleck: Acta Phys. Polon. B{\bf 26} (1995) 1869.
\bibitem{ssuzuki} S. Suzuki, O. Sakai and Y. Shimizu: Solid State Commun. {\bf 104} (1997) 429.
\bibitem{ludwig2} A. W. W. Ludwig: Int. J. Mod. Phys. {\bf B8} (1994) 347.
\end{thebibliography}
\end{document}